# Quantized conductance observed in reversible atomic contacts fabricated by template electroplating using an on-membrane anode

Zuxin Ye and Wenhao Wu

*Department of Physics, Texas A&M University, College Station, Texas 77843, USA*

We report a new template electroplating method for fabricating reversible atomic contacts between a single nanowire and a macroscopic contact pad. In comparison to typical template methods using a standing-alone anode, we directly evaporate the anode on one of the porous membrane surfaces. Single nanowires, upon emerging from the pores, make reversible atomic contacts with the on-membrane anode via a self-terminating mechanism. Quantized conductance steps have been observed in a controlled fashion during deposition and dissolution. This method can potentially be applied for the controlled fabrication and integration of nanowires, point contacts, and nanosized interconnects in template-based nanofabrication.

Quantum transport through nano or atomic constrictions has been studied extensively in recent years due to its important role in fundamental physics, as well as its potential applications in future electronic devices.[1-10] At room temperature, conductance quantization has been observed in narrow constrictions fabricated using various techniques, including controllable breaking junction,[1,2] atomic force microscopy,[3,4] scanning tunneling microscopy,[5] controllable contact of macroscopic wires,[6] and electrochemical deposition,[7-9] *etc*. In most of these approaches, an atomic constriction or gap was formed between pairs of identical macroscopic metallic electrodes. However, atomic constrictions between nanostructures and macroscopic electrodes might be of great importance. Due to the fact that future electronic devices may consist of many nano-scale elements, one of the critical challenges is to find a practical technique to connect these elements in a controlled fashion. Recently, Terabe *et al*.[10] have demonstrated a nanoscale mechanical switch that could replace semiconductor switches in future electronic devices by controlling the formation of an atomic bridge between two nanowires about 1 nm apart. However, their approach involves complicated nano-patterning processes. It is highly desirable to develop much simpler approaches that are able to implement similar functionality.

In this Letter, we report a new porous template-based electrochemical method for fabricating reversible atomic contacts between truly a single nanowire and a contact electrode of the different materials. Electroplating using porous membranes as templates has been adopted as a reliable and low cost method for fabricating large aspect ratio nanowires.[11, 12] Conventional template electroplating is carried out by coating one surface of a porous membrane with a thick metal film as the cathode (the back electrode) to form an array of nanowires in the pores. A standing-alone anode is usually placed in the electrolyte cell on the other side of the membrane. Recently, F. Elhoussine *et al.*[8] have demonstrated conductance quantization in the atomic contact between a single Ni nanowire and a macroscopic front electrode deposited on the surface of the membrane opposite to the cathode in a conventional template deposition process. The contact was believed to form between a Ni nanowire and a bulk Ni electrode. However, the exact nature of such contacts is not clear. Wu *et al.*[13] have investigated the structure of nanowires grown by an identical electroplating process using scanning electron microscopy (SEM). It was demonstrated clearly that when a nanowire emerged from a pore, a large mushroom head of the size of tens of microns formed. Therefore, the atomic contacts fabricated using conventional template electroplating methods are most likely formed between the large mushroom heads and the electrodes, rather than between nanowires and the electrodes. In our current study, the front electrode evaporated on the membrane surface opposite of the cathode is used as the anode for electroplating. The potential for electroplating is applied directly across this on-membrane anode and the back electrodes, as shown in Fig. 1. This new method limits electroplating to the very edge of the anode. It prevents mushroom formation and produces reversible atomic contacts between truly a single nanowire and the anode, as we will discuss below. Furthermore, with a micro-patterned anode, this method can potentially be applied for the controlled fabrication and integration of nanowires, point contacts, and nanosized interconnects in template nanofabrication.

Figure 1 schematically illustrates this new electrochemical method. Porous membranes of thickness about 60 μm and pore diameter about 20 nm, from Whatman International, Ltd., were used as the template for nanowire growth. Gold layers of 200 nm in thickness were thermally evaporated on opposite surfaces of the membrane to form the anode and the cathode. The layers are thick enough to completely block the pores in the covered areas. The large back electrode (the cathode) and the small front electrode (the anode) have a small overlap region. When a voltage is applied across the two electrodes, only pores right along the edge of the anode that overlaps with the cathode will have significant nanowire growth due to the steep electrical field distribution in this region. When a nanowire fills the pore and approaches the anode, the metal ions are guided by the local electric field and deposited onto the sharpest point of the nanowire as discussed in Ref. 9. This sharp electric field gradient guides the growth front toward to anode, thereby eliminates the formation of a mushroom head on the tip of a nanowire as it emerges from a pore. Although we have been able to routinely observe mushroom growth in conventional electrodeposition with a standing-alone anode, we have not been able to observe any mushroom formation with our new on-membrane anode approach. Consequently, the gap between the nanowire and the front electrode decreases to the atomic scale and eventually form a point contact between truly a nanowire and the anode. This contact shorts the front and the back electrodes and terminates nanowire growth, since most of the applied voltage is now across the current limiting resistor and the nanowire under the contact. In fact, with a low potential difference across the contact region, now the dissolution process dominates the reduction process and the atomic contact can break off. With a suitable choice of the applied bias voltage, $V_{bias}$, and the current-limiting resistor, $R_{lim}$, this "on" and "off" process can result in a narrow and oscillating atomic contact. We note that this new method eliminates electroplating on the on-membrane anode. Therefore, the atomic contact is made between a nanowire and a front electrode of the different materials. The voltages on $R_{lim}$ and across the membrane are measured by two voltmeters, V1 and V2, respectively, in Fig. 1. The sample current is determined from the measured voltage on $R_{lim}$. The electrolyte used for this study was a solution of 1 M/L $CoCl_2$ and 30 g/L Boric acid ($H_3BO_3$). Deposition was carried out with a $V_{bias}$ ranging from 0.1 to 2 V, depending on the contact conductance values we aim to produce. A smaller $V_{bias}$ or a larger $R_{lim}$ leads to a point contact with a smaller conductance value.

Robust single nanowire connection with the front electrode was observed reproducibly using a large $V_{bias}$ (larger than 1 V) and a small $R_{lim}$ (< 1 k$\Omega$). Fig. 2 shows the typical sample voltage (main frame) and current (inset) curves during a deposition process with $V_{bias}$ = 1.5 V and $R_{lim}$ = 500 $\Omega$. Before forming a single nanowire connection characterized by a large drop in voltage and rise in current shown in the right portion of the curves, the sample usually experiences several smaller voltage jumps, corresponding to the wetting of the pores and partial connection of the nanowire. The number of these smaller jumps varies among different samples. For example, in the sample shown in Fig. 2, the first voltage drop is most likely corresponding to a wetting process. The final resistance value of a single nanowire with a low resistance connection to the front electrode is consistent with the estimated value based on the bulk resistivity of Co.

Using a smaller $V_{bias}$ (< 1 V) and a proper $R_{lim}$ (~ 10 kΩ), one can obtain a quasi-balance between deposition and dissolution processes due to the self-terminating mechanism discussed earlier. Quantized conductance steps have been observed during either the deposition (Fig. 3a) or the dissolution process (Fig. 3b) using $V_{bias}$ = 0.5 V and $R_{lim}$ = 10 kΩ. As shown in Fig. 3a, during the deposition process, the conductance for this sample jumps between several rather stable plateaus that have the value of several quantum conductance units of $G_0$ before forming a stable connection with $12G_0$, where $G_0 = e^2/h$ is the conductance quantum with $e$ and $h$ being the electron charge and the Plank constant, respectively. The descending dissolution curve consists of a series of single $G_0$ steps corresponding to the disconnection of quantum conductance channels, as indicated by the arrows in Fig. 3b. In Fig. 4, we plot for this sample a histogram of the conductance measured near the threshold of contacting. The conductance was measured every 0.05 second, and the total time for collecting the data in Fig. 4 was ~ 10 minutes. The bin size of the histogram was 0.1 $G_0$. Here we observe well-defined peaks in the histogram, with a nearly equal spacing in conductance values. Similar conductance histogram was observed in all the samples (more than 20 samples) with similar values of $V_{bias}$ and $R_{lim}$. Note that this spacing is not exactly $G_0$ due to multiple effects. Firstly, the electrolyte adds a conduction channel that is parallel to the nanowire-anode contact. The effective DC resistance of the electrolyte over a spacing of 1 mm is ~ 10 kΩ, which is of the order $G_0$. However, the question of how the electrolyte conduction channel and the structure arrangement of the atomic contact affect each other is beyond the scope of the current study. Secondly, voltage measurement is made between the anode and the cathode. This includes the voltage drop along the nanowire and the voltage across the atomic contact. The calculated conductance for the point contact is given by $G = 1/(R_w + R_c) = G_c/(1 + G_c/G_w)$, where $R_w$ and $R_c$ are the resistance values for the nanowire and the nano constraint, respectively, and $G_w$ and $G_c$ are the conductance values for the nanowire and the nano constraint, respectively. Since $R_w$ is ~ 2-3 k$\Omega$ for Co wires of 20 nm in diameter and 60 μm in length, the measured conductance steps is renormalized by a factor of $(1 + G_c/G_w)$. This effect possibly explains why the conductance spacing in the histogram is smaller than the conductance quantum $G_0$. We note that a similar effect related to a limiting resistor in series connection with a point contact was studied by Boussaad and N. J. Tao.[9] Another effect could lead to the smaller-than-$G_0$ conductance step is reflection in the conducting channels of the point contact.[14] Nevertheless, the histogram in Fig. 4 clearly demonstrates the quantized conductance steps as conductance channels are added to or removed from a narrow constraint. The observed conductance steps are consistent with $e^2/h$, which is expected for ferromagnetic Co nanowires. In Fig. 5a we show a stable contact oscillating about $6G_0$. With appropriate settings for $V_{bias}$ and $R_{lim}$, it is also possible to observe conductance oscillations between 0 and $G_0$, behaving as if it is a two-level system, as shown in Fig. 5b.

We note that the nature of the atomic contact described here is different from those described in earlier publications based on conventional template electroplating methods. Wu et al.[13] have shown in their SEM studies of single-contact nanowires that large mushroom heads grow on the tips of nanowires as they emerge from the pores using the conventional template methods. Mushroom heads form since nanowires emerging from the pores are usually located at distances from the front electrode that are far larger than the diameter of the pores. Therefore, contacts in those systems are made between such large mushroom heads and bulk electrodes. Our new method with an on-membrane anode forms atomic contacts between nanowires and bulk electrodes without producing mushroom heads at the tips of nanowires. Our SEM studies of samples fabricated by this new method have not observed any mushroom formation. This unique feature allows the investigation of ballistic transport between a nanowire and a macroscopic contact of the different materials.

In summary, we have developed a new template-based electrochemical method with an on-membrane anode. We have demonstrated that reversible atomic contacts between truly single nanowires and bulk metals of different materials can be fabricated based on a self-terminating mechanism. Quantized conductance steps and oscillations have been observed during controlled fabrication of such atomic contacts. Electrochemical growth methods similar to the one described here may provide efficient and low-cost techniques for fabricating and integrating nanosized elements and interconnects using porous membrane templates. This work was supported by NSF Grant No. DMR-0551813.

**FIGURE CAPTIONS**

Fig. 1  Schematic drawing of the experimental setup. The front and the back electrodes serve as the anode and the cathode, respectively. Voltmeter V1 measures the potential drop across a current limiting resistor $R_{lim}$, from which the current is determined. Voltmeter V2 measures the potential difference between the cathode and the anode. Only pores right along the edge of the front electrode have significant nanowire growth due to the high local electrical field near the edge.

Fig. 2  Time evolution of sample voltage (main frame) and current (inset) during Co nanowire growth. The first drop in the voltage (increase in the current) is due to the wetting of the pores. The second one indicates contact formation between a single nanowire and the front electrode.

Fig. 3  Conductance in units of $e^2/h$ for a sample during (a) deposition and (b) dissolution. During deposition, jumps of a few units of $e^2/h$ between plateaus are observed. During dissolution, stepwise decreases in the conductance, corresponding to the disconnection of quantum conductance channels, are observed, as indicated by the arrows.

Fig. 4  Conductance histogram for a contact between a Co nanowire and an Au electrode demonstrating quantized conductance in the point contact.

Fig. 5  Conductance measured on (a) a stable atomic contact and (b) an unstable atomic contact with its conductance oscillating in a way similar to that of two-level systems.

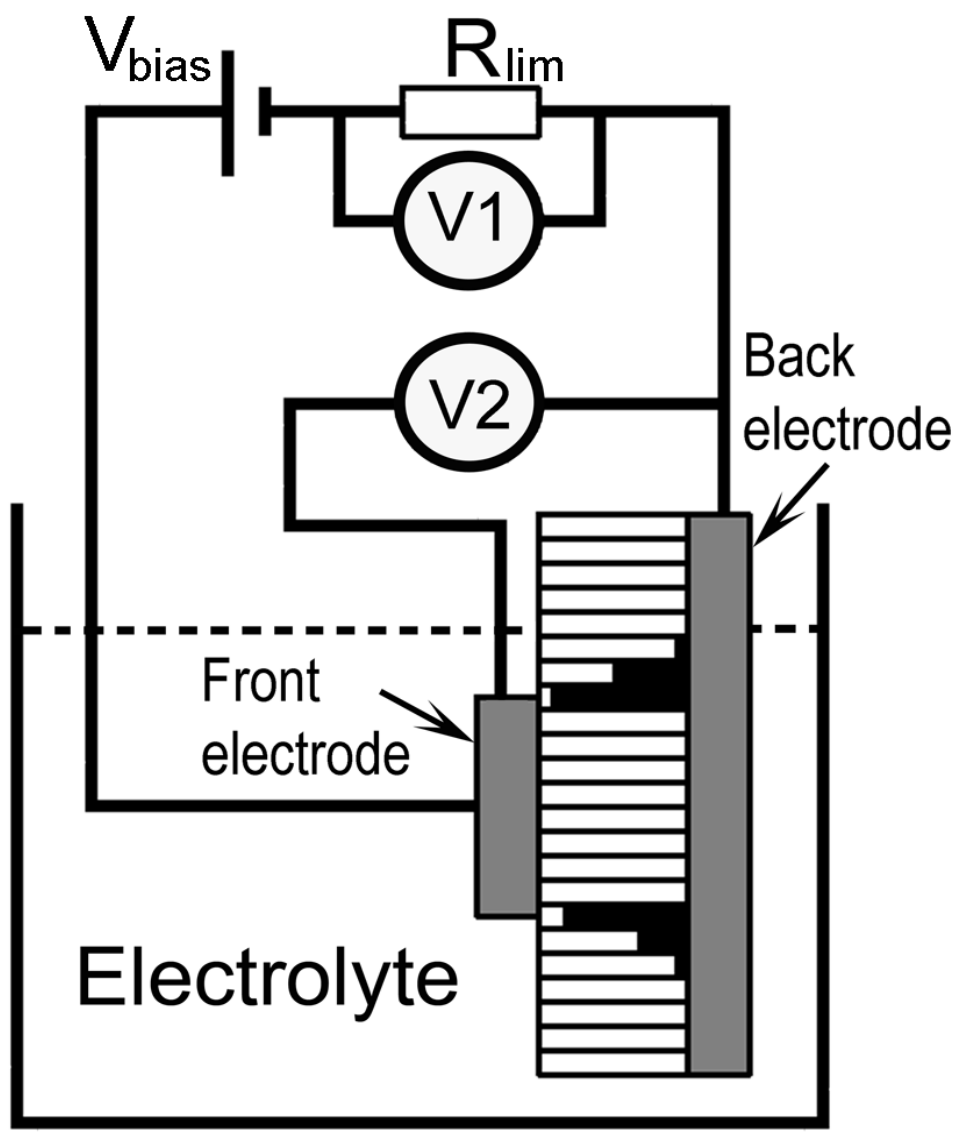

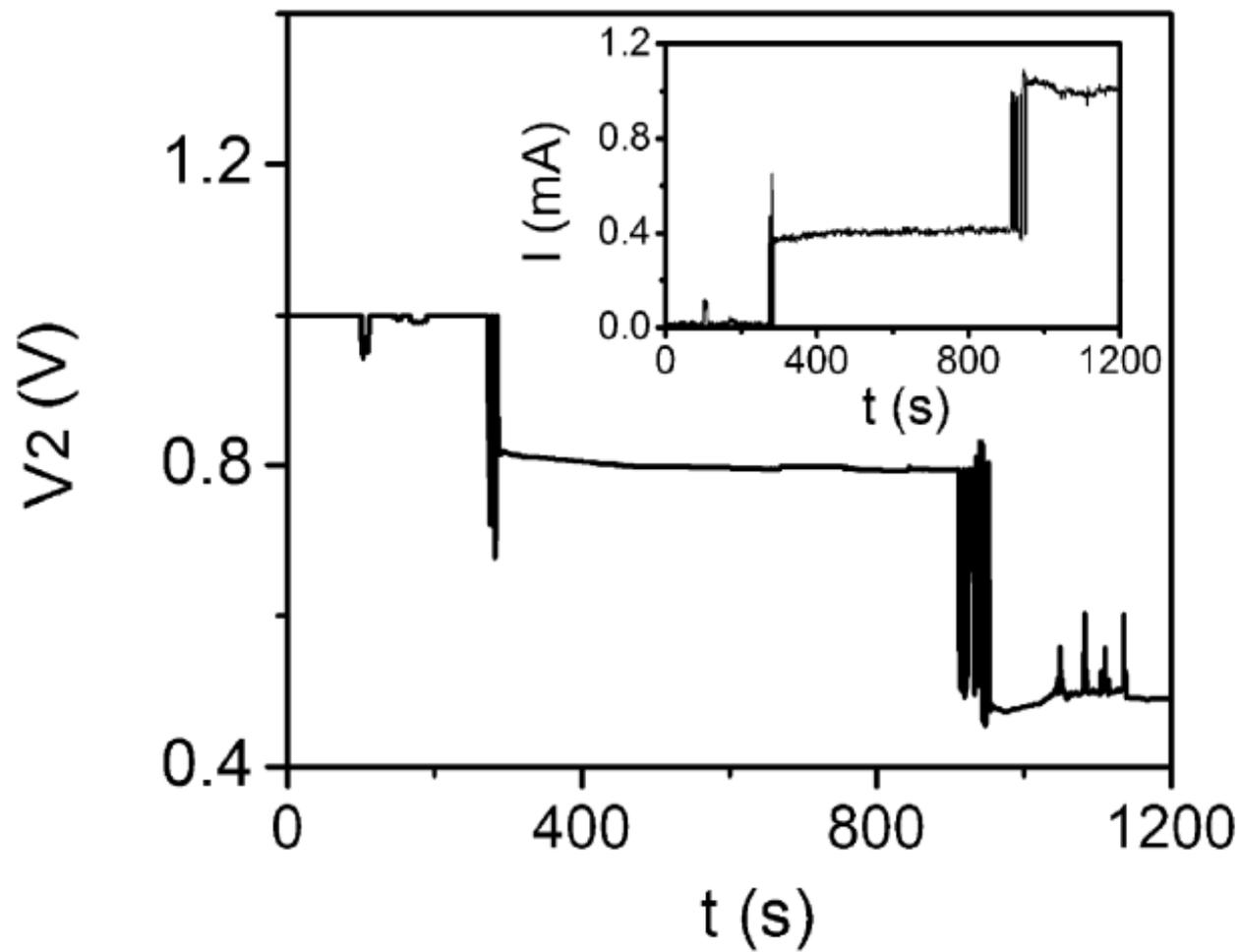

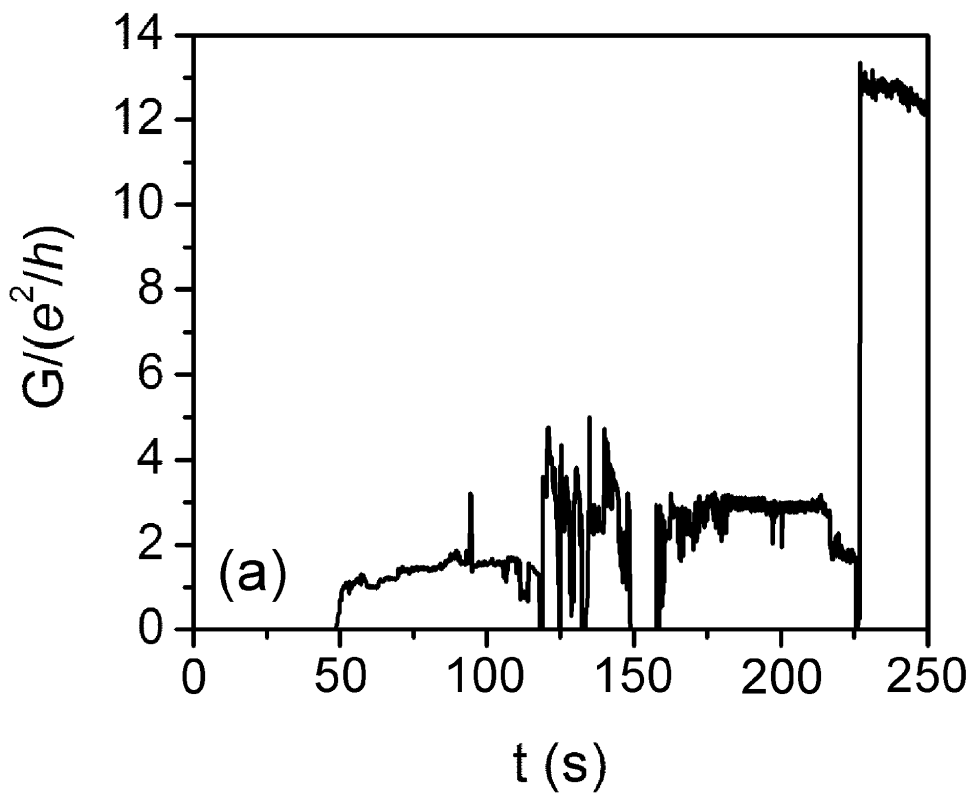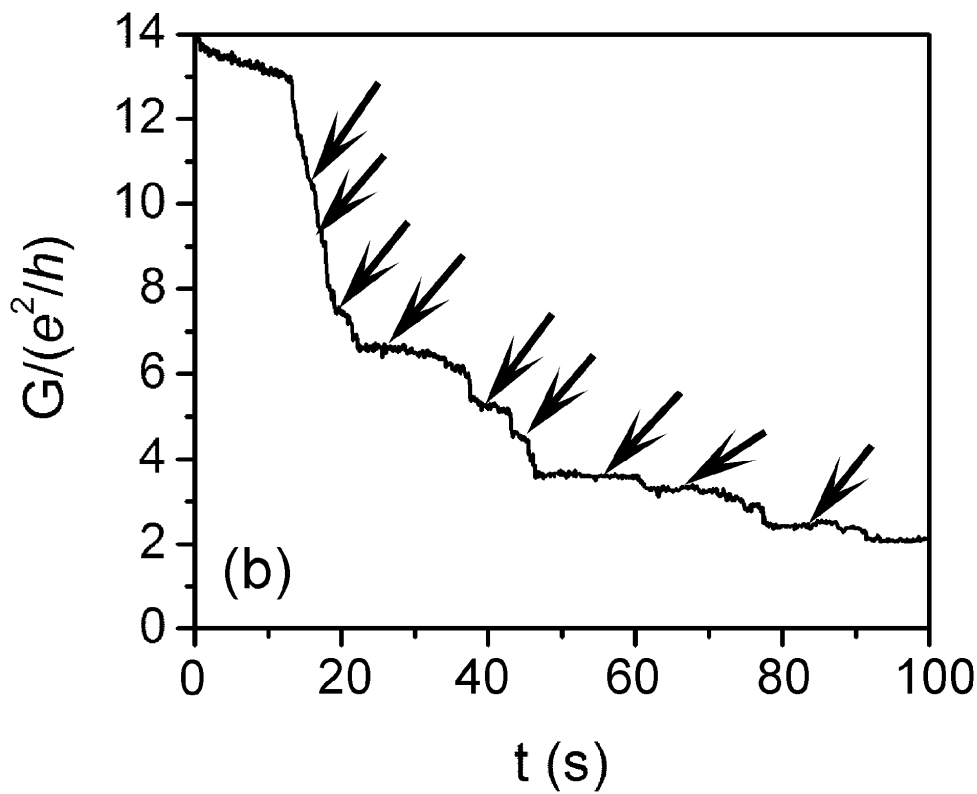

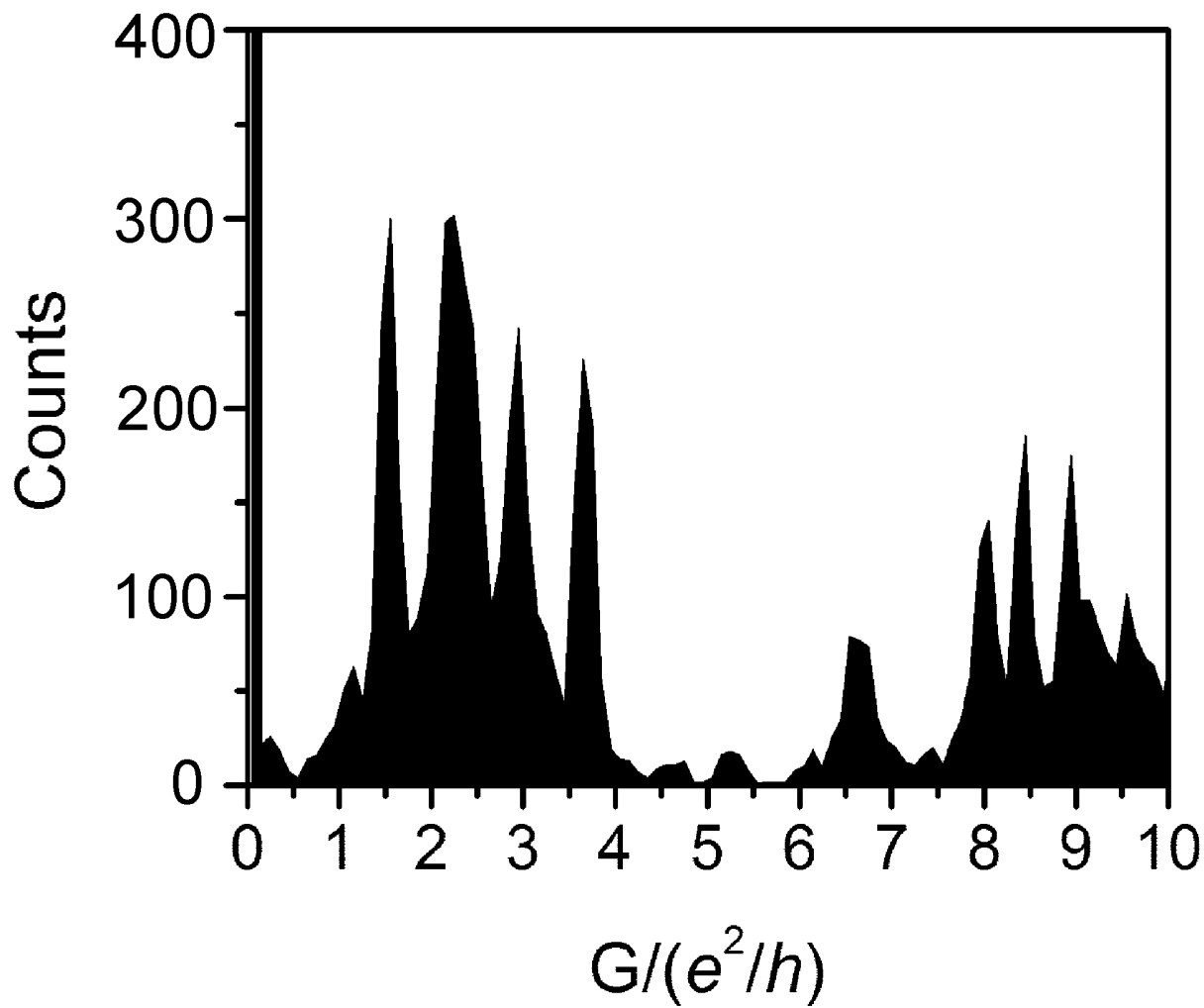

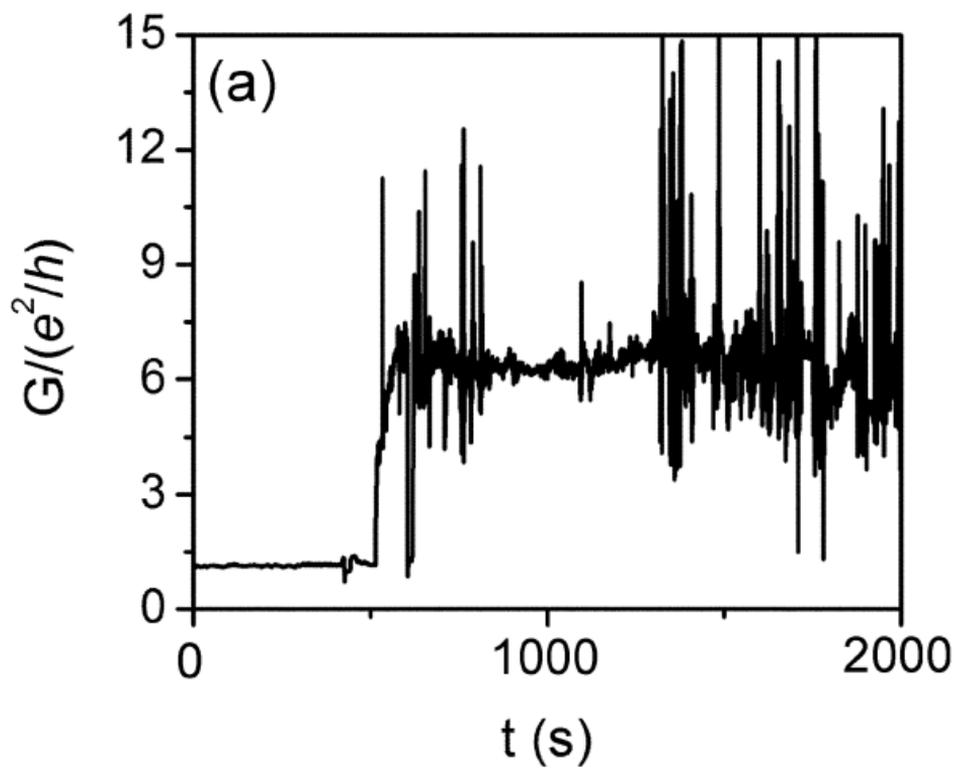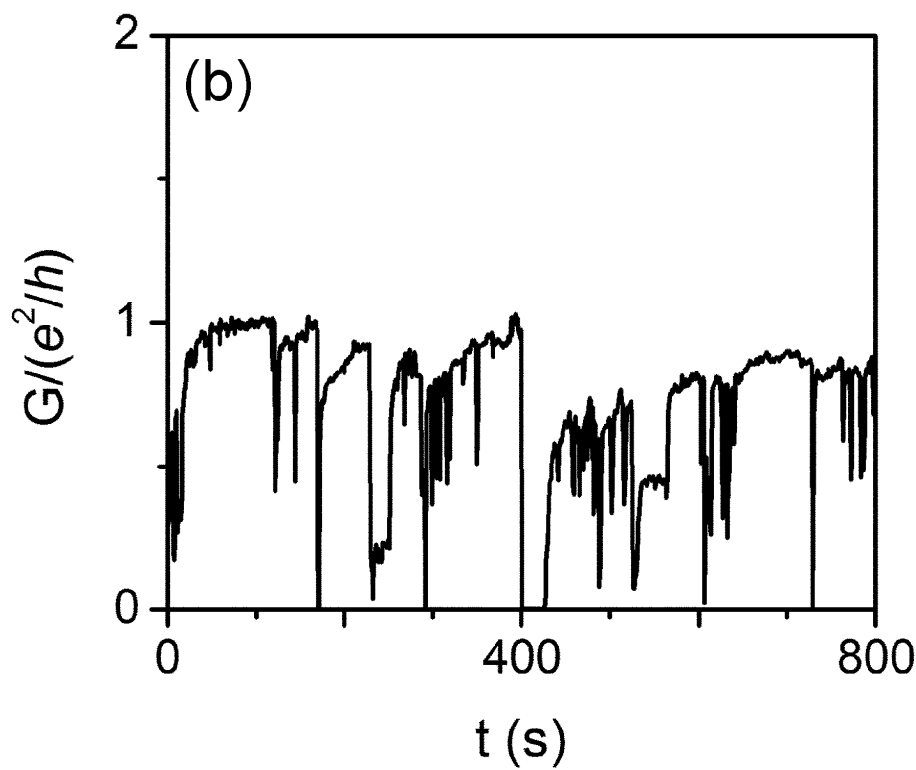